\documentclass[aoas,dvips]{arximspdf}

\doi{10.1214/10-AOAS387}
\volume{4}
\issue{4}
\pubyear{2010}
\firstpage{1642}
\lastpage{1643}

\begin{document}
\begin{frontmatter}

\title{Leo Breiman}
\runtitle{Leo Breiman}

\begin{aug}
\author[A]{\fnms{Michael I.} \snm{Jordan}\ead[label=e1]{jordan@stat.berkeley.edu}\corref{}}
\runauthor{M. I. Jordan}
\affiliation{University of California, Berkeley}
\address[A]{Department of Statistics\\
University of California, Berkeley\\
Berkeley, California 94720\\
USA\\
\printead{e1}} 
\end{aug}

\received{\smonth{7} \syear{2010}}



\end{frontmatter}

Statistics is a uniquely difficult field to convey to the uninitiated.
It sits astride the abstract and the concrete, the theoretical and the
applied. It has a mathematical flavor and yet it is not simply a branch
of mathematics.  Its core problems blend into those of the disciplines
that probe into the nature of intelligence and thought, in particular
philosophy, psychology and artificial intelligence.  Debates over
foundational issues have waxed and waned, but the field has not yet
arrived at a single foundational perspective.

Given these complexities it might seem surprising that human beings
could have definite opinions about core issues in statistics, and
surprising that working in such a field could be pleasurable.  And
yet there was Leo Breiman, who had his definite opinions about statistics
and who took great pleasure in waking up every morning to see what more he
could do to bring the field along.

To the extent that most statisticians have a vision about the
final conclusive form the field might take, I suspect that this
vision is a mathematical one---a set of core definitions, axioms
and theorems.  Moreover, I think that many statisticians will
expect for these mathematical ideas to involve a set of
optimality principles, such that it will be possible for a
user of statistics circa 2500 AD to dial in the description of
a problem and out will pop the optimal procedure.

I think that Leo had come to a different vision.  In thinking
about Leo I think about the box of tools in my basement.  It
contains hammers, screwdrivers, pliers, nails, screws and rivets.
Of the infinite number of possible physical forms that objects
for manipulating the physical world could have taken, these are
the ones that have come to us from our ancestors in the applied
field of ``management of uncertainty in physical structures.''
They arose via little bits of human genius and they have stood
the test of time.

My vision of Leo's vision is, of course, an inference, and to support
my inference I will exhibit some of the (anecdotal) data on which
it is based.

\begin{itemize}
\item I first met Leo at a conference where I found myself at a
lunch table with Leo and Jerry Friedman.  Leo initiated the
lunchtime conversation as follows:  ``Jerry, what do you think
about nearest neighbor?''  An outsider might have naively thought
that ``nearest neighbor'' was some recent fad, but of course Leo
(and Jerry) had been thinking about nearest neighbor for decades.
Leo simply wanted to kick nearest neighbor around the block yet
again, testing it against recent developments (e.g., hardware,
algorithmic and mathematical) to see whether new reasons might
have emerged for preferring nearest neighbor as a solution for
certain kinds of applied problems.  In subsequent years I had
this kind of conversation with Leo many times, and it left a strong
sense that Leo was thinking carefully about the ``ultimate toolbox.''
(He did not seem enthralled by nearest neighbor, but he did seem
to think that the jury was still out.)

\item Consistent with the notion of constructing a toolbox, Leo
delighted in terminology for statistical procedures that have a
physical connotation.  Some of these he invented, others he borrowed.
Consider the following examples: ``curds and whey,'' ``bagging,''
``stacking,'' ``garotte,'' ``CART'' and ``forests.''  One seeming
counterexample is ``ACE,'' but I have had the advantage of seeing
Jerry Friedman's tribute to Leo, where I learned that ``ACE'' was
Jerry's invention.  Moreover, Jerry relates that Leo only went
along with ``ACE'' when he realized that it is the name of a hardware
store.  Case closed!

\item Leo was an artist and I think that he would have subscribed to
a vision in which the users of statistics in 2500 AD would be given
a set of tools in which they could express their own creative
solutions to problems.  He would have rejected a rigid prescription
of what must be done in each and every situation.

\item Another preferred piece of Breimanesque terminology was
``off-the-shelf,'' again a rather physical metaphor.  Leo tended
to be suspicious of ``free parameters''; procedures should work
with little or no ``tuning.''  Whenever I might tell Leo about
a project of mine that had a Bayesian flavor this suspicion would
come to the fore.  While this might seem inconsistent with the
notion of the ``statistician as artist,'' the more refined notion
is perhaps that of the ``statistician as artisan.'' Hammers and
screwdrivers don't tend to have free parameters.  Leo wanted each
of the basic tools to do one job and do it well.

\end{itemize}

Leo was a complex human being, and my portrait may at best have captured
only one aspect of the man.  But as Leo taught us, ensembles often do
surprisingly well, and hopefully the ensemble of portraits will do
him justice.

\printaddresses

\end{document}